\documentclass[aps,prl,twocolumn,floatfix,showpacs,groupaddress]{revtex4-2}

\usepackage{amssymb}
\usepackage{amsmath}
\usepackage{bm}
\usepackage{graphicx}
\usepackage{hyperref}
\usepackage{xcolor}
\usepackage{soul}

\usepackage[utf8]{inputenc}

\begin{document}

\title{Friction mediated phase transition in confined active nematics}

\author{Cody D. Schimming}
\email[]{cschim@lanl.gov}
\affiliation{Theoretical Division and Center for Nonlinear Studies, Los Alamos National Laboratory, Los Alamos, New Mexico, 87545, USA}

\author{C. J. O. Reichhardt}
\affiliation{Theoretical Division and Center for Nonlinear Studies, Los Alamos National Laboratory, Los Alamos, New Mexico, 87545, USA}

\author{C. Reichhardt}
\affiliation{Theoretical Division and Center for Nonlinear Studies, Los Alamos National Laboratory, Los Alamos, New Mexico, 87545, USA}

\begin{abstract}
Using a minimal continuum model, we investigate the interplay between circular confinement and substrate friction in active nematics. Upon increasing the friction from low to high, we observe a dynamical phase transition from a circulating flow phase to an anisotropic flow phase in which the flow tends to align perpendicular to the nematic director at the boundary. We demonstrate that both the flow structure and dynamic correlations in the latter phase differ from those of an unconfined, active turbulent system and may be controlled by the prescribed nematic boundary conditions. Our results show that substrate friction and geometric confinement act as valuable control parameters in active nematics.
\end{abstract}

\maketitle

A remarkable feature of active fluids is their ability to generate macroscopic flows from energy consumption at the micro-scale \cite{ramaswamy10,marchetti13}. In many cases, however, these flows are chaotic, a phenomenon dubbed ``active turbulence'' due to its qualitative similarities to inertial turbulence \cite{wensink12,sanchez12,creppy15,nishiguchi15,yang16,genkin17,doo17,alert20,carenza20,alert22}. Identifying methods to control the flows generated by active fluids has recently been of particular interest due to potential technical and biomedical applications. Efforts in this direction have included coupling to concentration gradients, patterning activity, manipulating sample geometry, and imposing boundary conditions \cite{guillamat16,zhou17b,sokolov19,hardouin20,koizumi20,norton20,reichhardt20,thijssen21,vinze21,zhang21b,figueroa22}. Here, we focus on ``active nematics,'' active fluids composed of elongated constituents that produce macroscopic flows via force dipoles, and study the flow patterns that emerge from the interplay between two important and relevant control mechanisms: circular confinement \cite{wioland13,guillamat16b,theillard17,wu17,norton18,opathalage19,hardouin20,norton20,mozaffari21,guillamat22} and substrate friction \cite{doo16,doo16c,putzig16,sriv16,doo19,thijssen20,thijssen20b}. While the effects of these control mechanisms on the dynamical behavior of active nematics have been previously studied independently, their interplay has remained unexplored. Energy dissipation through frictional damping introduces a length scale, the hydrodynamic screening length, which sets the scale below which hydrodynamic interactions are important. Further, because nematics are inherently anisotropic, confinement allows the prescription of topologically and geometrically distinct boundary conditions. While it is known that the boundary conditions do not alter the flow state for frictionless systems \cite{norton18}, it is not known whether a paradigm exists in which the boundary conditions can tune the system dynamics.  

Here, using a minimal continuum model, we show that when the hydrodynamic screening length is decreased, circularly confined active nematics transition from a circulating flow state to a dynamical anisotropic flow phase that, to our knowledge, has not been previously described. We show that the anisotropic flow phase is distinct from active turbulence, and is characterized by flows and vortices that organize perpendicular to the nematic boundary condition. As a result, the boundary conditions may be used to tune the dynamics and correlation timescales of the system. To investigate the interplay between confinement and hydrodynamic screening, we vary the screening length at a fixed average time scale associated with active stress injection \cite{giomi14}. This differs from previous investigations of the effects of substrate friction on bulk active nematics in which only the time scale associated with viscous forces is fixed, while the time scale associated with frictional damping is increased leading to the suppression of flow \cite{thampi14,doo16,doo16c,thijssen20}. We find that the anisotropic flow transition occurs when the elastic interactions between defects become dominant due to the hydrodynamic screening length dropping below the size of the topological defects. Our results not only shed light on how biological systems, which tend to have larger screening lengths, organize flow and dynamics, but also can be used to engineer controlled flow and dynamics by employing hydrodynamic screening and boundary conditions as control parameters. 

The numerical model we use has been previously well documented \cite{marenduzzo07,doo18}. We briefly review it here and give specific details in the Supplementary Material \cite{SuppNote23}. The equations for the active nematic are written in terms of the nematic tensor order parameter $\mathbf{Q}$, the fluid velocity $\mathbf{v}$, and the fluid pressure $p$:
\begin{gather} 
    \frac{\partial \mathbf{Q}}{\partial t} + (\mathbf{v} \cdot \nabla) \mathbf{Q} - \mathbf{S} = -\frac{1}{\gamma}\frac{\delta F}{\delta \mathbf{Q}}, \label{eqn:QEvo} \\
    -\eta \nabla^2 \mathbf{v} + \Gamma \mathbf{v} = -\nabla p - \alpha \nabla \cdot \mathbf{Q}, \quad \nabla \cdot \mathbf{v} = 0. \label{eqn:Velocity}
\end{gather}
Equation \eqref{eqn:QEvo} describes the time evolution of the nematic tensor order parameter $\mathbf{Q} = S\left[\mathbf{n} \otimes \mathbf{n} - (1/2) \mathbf{I}\right]$ where $S$ gives the local degree of order and $\mathbf{n}$ is the nematic director. $\mathbf{S} = \mathbf{S}(\mathbf{Q},\nabla \mathbf{v})$ is a generalized tensor advection \cite{beris94}, $F$ is the usual Landau-de Gennes free energy in which we assume one-constant elasticity \cite{deGennes75}, and $\gamma$ is a rotational viscosity. Equation \eqref{eqn:Velocity} is the modified Stokes equation describing low Reynolds number flows. Here $\eta$ is the fluid viscosity. The terms proportional to $\Gamma$ and $\alpha$ are additions to the usual Stokes equation and they describe, respectively, friction between the active nematic and substrate and the strength of active forces in the nematic \cite{doo18}. $\alpha > 0$ corresponds to extensile forces while $\alpha < 0$ corresponds to contractile forces. The divergence free condition on the velocity models an incompressible fluid. A discussion of the length and time scales associated with the model is given in the Supplementary Material \cite{SuppNote23}.

We non-dimensionalize, discretize, and solve Eqs. \eqref{eqn:QEvo} and \eqref{eqn:Velocity} numerically on a circular domain using fixed nematic boundary conditions with the Matlab/C++ finite element package FELICITY \cite{walker18}. We fix the domain radius to $\tilde{R} = 7.5$ in dimensionless units (see Supplementary Material \cite{SuppNote23} for details on dimensionless quantities). We also fix the ratio $\tilde{\alpha}/(\tilde{\eta} + \tilde{\Gamma}) = 1$ and vary only the hydrodynamic screening length $\tilde{L}_{SC} = \sqrt{\tilde{\eta}/\tilde{\Gamma}}$. The tildes denote dimensionless quantities and are omitted in what follows for brevity. This procedure differs from previous explorations of the effect of friction on bulk active nematics in that we do not hold the viscosity constant \cite{thampi14,doo16,doo16c,doo19,thijssen20}. As detailed in the Supplementary Material, constraining the ratio $\alpha/(\eta + \Gamma) = 1$ fixes the average active time scale associated with viscous and frictional forces and yields a dimensionless Stokes equation in which the active force is relevant across all scales of the hydrodynamic screening length. This allows us to isolate the effect of the screening length without suppressing the flows generated by the active stress \cite{SuppNote23}. We choose $\alpha/(\eta + \Gamma) = 1$ since this leads to the circulation phase in the zero friction limit for our chosen domain size \cite{norton18}. Much smaller values lead to a quiescent state, while much larger values lead to turbulent behavior.

\begin{figure}
    \centering
    \includegraphics[width = \columnwidth]{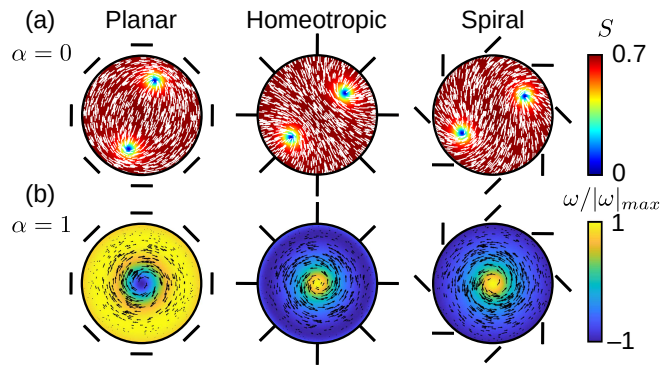}
    \caption{(a) Non-active ($\alpha = 0$) nematic configurations for the three boundary conditions studied: planar, homeotropic, and spiral anchoring. The color in the plots shows the local scalar order parameter $S$ while the white lines show the nematic director $\mathbf{n}$. Lines outside the domain depict the fixed orientation of the nematic director at the boundary. (b) Example velocity and vorticity fields for simulated active ($\alpha = 1$) nematics with $L_{SC} = 10$. The color shows the normalized vorticity field, while the arrows show the magnitude and direction of the velocity.}
    \label{fig:CirculationVorticities}
\end{figure}

We consider three nematic boundary conditions: planar, homeotropic, and spiral. Figure \ref{fig:CirculationVorticities}(a) shows the non-active ($\alpha = 0$) state for each of these boundary conditions. All three boundary conditions impose an overall topological charge of $+1$ on the system, so topological defects (points of singular nematic orientation, called disclinations) must form. In the non-active state, the lowest energy configuration consists of two $+1/2$ winding number disclinations that lie on opposite ends of the domain. In active nematics, $+1/2$ disclinations are motile, and so the configurations we consider show dynamical behavior at lower activities than bulk systems with zero overall topological charge \cite{doo18}.

For the active system ($\alpha = 1$), varying $L_{SC}$ induces a clear transition between two distinct dynamical phases. For large $L_{SC}$ (low friction) the long-time dynamical behavior of the system is characterized by circulating flow. For small $L_{SC}$ (high friction), the circulation ceases and a dynamical anisotropic flow phase reminiscent of active turbulence emerges. Unlike traditional active turbulence, the anisotropic flow is characterized by long, thin vortices that organize near the boundary to lie perpendicular to the nematic director.

The circulation phase observed at large $L_{SC}$ is depicted in Fig. \ref{fig:CirculationVorticities}(b), where we plot the velocity and vorticity fields for the three boundary conditions at $L_{SC} = 10$. In all cases, a central vortex is formed and the flow circulates in a clockwise or counter-clockwise direction. For the level of activity we consider, the nematic configuration initially contains two $+1/2$ defects circulating each other at early times  that eventually merge causing the configuration to develop a central $+1$ defect with a spiral pattern for all boundary conditions (Fig. S1). 

The direction of circulation is a spontaneously broken symmetry for planar and homeotropic anchoring, since these boundary conditions are achiral; however, the spiral boundary conditions break chiral symmetry and always produce counter-clockwise flow. If the boundary conditions were rotated by $\pi/2$, the resulting flow would circulate in the opposite direction. Hence, the spiral boundary condition offers a method of controlling the direction of flow, similar to that shown in experiments with bacterial suspensions in a pre-patterned liquid crystal \cite{koizumi20}  except that it is not necessary to pre-pattern the entire liquid crystal, but only the director at the boundary of the sample. 

\begin{figure}
    \centering
    \includegraphics[width = \columnwidth]{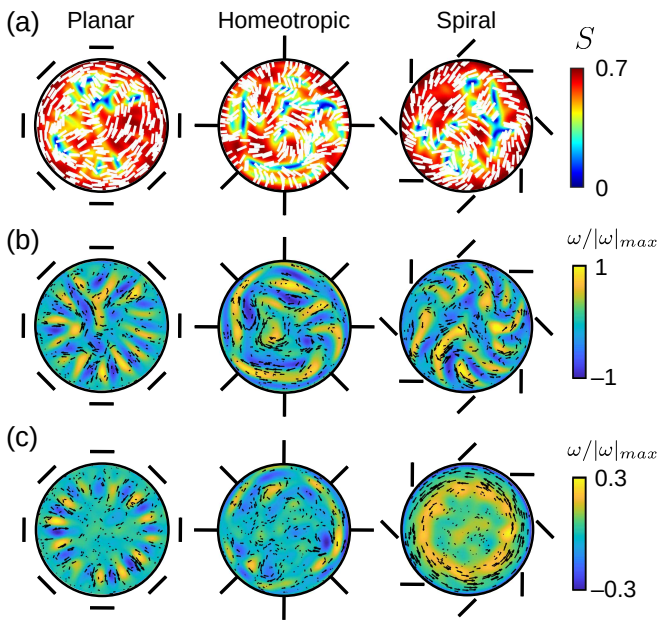}
    \caption{(a) Time snapshots of the nematic configurations for simulated active nematics with $L_{SC} = 0.2$. (b) Velocity and vorticity fields at same time snapshot as in (a). (c) Time averaged velocity and vorticity fields for the same simulations.}
    \label{fig:PerpVorticities}
\end{figure}

In contrast, the dynamics of the anisotropic flow phase at small $L_{SC}$ depend on the choice of nematic boundary condition. Figures \ref{fig:PerpVorticities}(a) and \ref{fig:PerpVorticities}(b) show time snapshots of the nematic configuration and velocity and vorticity fields for each boundary condition at $L_{SC} = 0.2$. The many long, thin vortices in this phase tend to lie perpendicular to the fixed nematic director at the boundary, and as a result, the flow direction is influenced by the prescribed boundary conditions. 

 While the anisotropic flow phase is qualitatively reminiscent of traditional active turbulence, we show in Fig.~\ref{fig:PerpVorticities}(c) that the time-averaged velocity and vorticity fields retain structure when averaged over the length of the simulation. This differs from the zero flow time average obtained in a chaotic, turbulent system, as seen in simulations of unconfined active nematics with periodic boundary conditions (Fig. S2). As shown in Fig.~\ref{fig:PerpVorticities}(c), for both planar and homeotropic boundary conditions we find persistent organization of vortices near the boundary. The spiral boundary conditions produce time-averaged circulating flow near the boundary instead of the distinct spiral vortex pattern found in the time snapshot of Fig.~\ref{fig:PerpVorticities}(b). This is because the dynamics of the vortices are relatively static for planar and homeotropic boundary conditions, but circulate for spiral conditions as a result of the promotion of circulating flows (see Supplementary Movies 1--3).

The primary mechanism behind the perpendicular alignment of the flow field to the nematic director at the boundary is the active nematic bend instability \cite{ramaswamy07}, which promotes undulations in the nematic director that form parallel to the director. Since $L_{SC}$ controls the size of vortices \cite{doo16}, it also controls the size of the undulations. When $L_{SC}$ is small enough, the undulations become large enough to support the unbinding of $\pm 1/2$ disclination pairs that generate flows perpendicular to the director. Thus, the nematic configuration in the anisotropic flow phase is characterized by motile $+1/2$ disclinations unbinding near the boundary and then, at a later time, annihilating with immotile $-1/2$ disclinations that remain near the boundary (see Supplementary Movies 1--3).

To quantitatively describe the system, we define two parameters related to the velocity of the fluid. The circulation parameter is \cite{opathalage19},
\begin{equation} \label{eqn:Circulation}
    \Phi = \left\langle \frac{v_{\theta}}{|\mathbf{v}|}\right\rangle
\end{equation}
where $v_{\theta}$ is the azimuthal component of the velocity. All averages are computed over the full simulation time and spatial domain. For coherent circular flows, $\Phi = \pm 1$, while for chaotic, active turbulent flows, $\Phi = 0$. We also measure the average ratio of flow perpendicular to the nematic director boundary condition $\mathbf{n}_0$ to that parallel to $\mathbf{n}_0$:
\begin{equation} \label{eqn:PerpFlow}
    v_{\perp} = \left\langle \frac{|\mathbf{v}\times \mathbf{n}_0|}{|\mathbf{v}\cdot\mathbf{n}_0|} \right\rangle.
\end{equation}
We note that this perpendicular flow measure depends on the boundary condition. For a chaotic, active turbulent state we expect $v_{\perp} = 1$, that is, an equal proportion of perpendicular and parallel flows.

\begin{figure}
    \centering
    \includegraphics[width = \columnwidth]{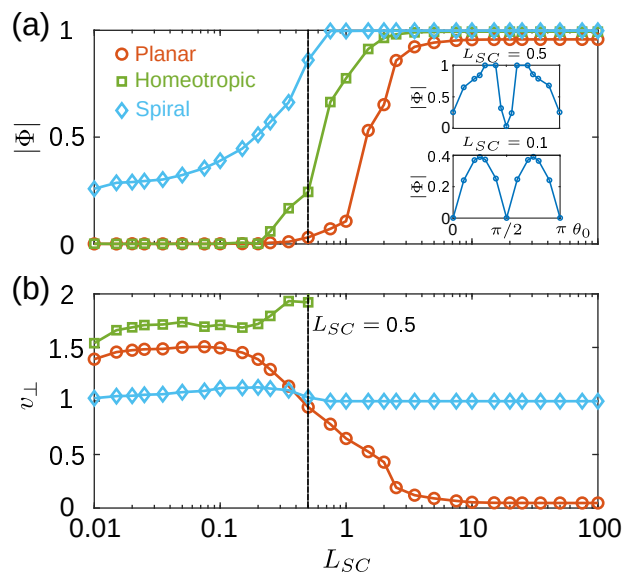}
    \caption{(a) Flow circulation $|\Phi|$ vs hydrodynamic screening length $L_{SC}$ for confined active nematic systems with planar, homeotropic, and spiral anchoring. Inset: $|\Phi|$ vs anchoring angle $\theta_0$ for $L_{SC} = 0.5$ and $L_{SC} = 0.1$. (b) Perpendicular flow parameter $v_{\perp}$ vs $L_{SC}$. The dashed line in both plots marks $L_{SC} = 0.5$, below which the confined system is in the anisotropic flow phase.}
    \label{fig:FlowMeasures}
\end{figure}

Figure \ref{fig:FlowMeasures} shows $|\Phi|$ and $v_{\perp}$ versus $L_{SC}$ for systems with hydrodynamic screening ranging over several orders of magnitude. For planar and homeotropic anchoring, the circulation parameter $|\Phi|$ ranges from $1$ at large $L_{SC}$ to $0$ at small $L_{SC}$. The spiral boundary conditions always have nonzero circulation, but $|\Phi|$ decreases as $L_{SC}$ is reduced. We also measured $|\Phi|$ for systems with anchoring angle $\theta_0 \in [0,\pi]$ with respect to the boundary and $L_{SC} = 0.5$ and $L_{SC} = 0.1$, as shown in the inset of Fig. \ref{fig:FlowMeasures}(a). We find that in the anisotropic flow phase ($L_{SC} = 0.1$) the maximal circulation occurs for spiral anchoring, $\theta_0 = \pi/4, 3\pi/4$. However, near the phase transition ($L_{SC} = 0.5$) the circulation is maximal close to planar anchoring, $\theta_0 = \pi/2$, before dropping to zero exactly at planar anchoring. In this case, the circulating phase is stabilized by the shear flow resulting from the central vortex shown in Fig. \ref{fig:CirculationVorticities}(b). The nematic tends to align at an angle relative to the flow, the ``Leslie angle,'' which for our system is close to planar anchoring \cite{leslie92,SuppNote23}. Thus, if $\theta_0$ is close to the Leslie angle, the circulating phase will remain stable for a larger range of $L_{SC}$. This indicates that the transition itself may be tunable by material parameters that control the flow alignment.

While $|\Phi|$ serves as an order parameter that indicates a transition between dynamical phases, the perpendicular flow parameter $v_{\perp}$ quantifies the nature of the anisotropic flow phase at small $L_{SC}$. Since $v_{\perp}$ depends on the director $\mathbf{n}_0$ at the boundary, its definition changes for different boundary conditions. For example, for planar boundary conditions we obtain $v_{\perp} = \left\langle |v_r|/|v_{\theta}|\right\rangle$, where $v_r$ is the radial component of the velocity, while for homeotropic boundary conditions we obtain the reciprocal, $v_{\perp} = \left \langle |v_{\theta}|/|v_r| \right\rangle$. For coherent circulating flows, then, $v_{\perp}$ goes to zero for planar boundary conditions but diverges for homeotropic boundary conditions. Remarkably, in the anisotropic flow phase $v_{\perp}$ is very similar for the planar and homeotropic cases, even though $v_{\perp}$ is defined reciprocally. We mark the transition to anisotropic flow in Fig. \ref{fig:FlowMeasures} as occurring at $L_{SC} = 0.5$, since below this value, $v_{\perp} > 1$ for the three considered boundary conditions. For our choice of model parameters \cite{SuppNote23}, $L_{SC} = 0.5$ is roughly the radius of the topological defects, suggesting that the hydrodynamic interaction between defects promotes circulation, and that the transition to anisotropic flow occurs when elastic interactions (i.e., Coulomb-like interactions) between defects become dominant. 

\begin{figure}
    \centering
    \includegraphics[width = \columnwidth]{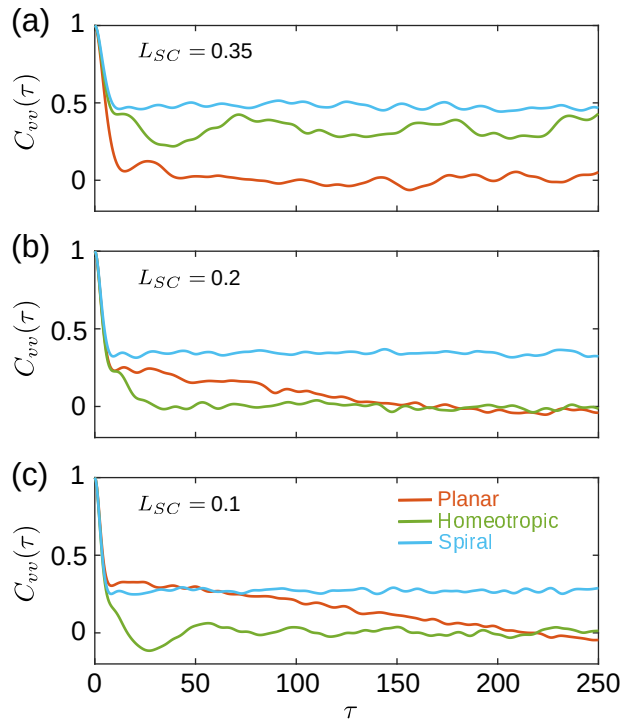}
    \caption{Velocity time correlation function $C_{vv}(\tau)$ plotted for simulations with hydrodynamic screening lengths (a) $L_{SC} = 0.35$, (b) $L_{SC} = 0.2$, and (c) $L_{SC} = 0.1$ for planar, homeotropic, and spiral anchoring.}
    \label{fig:CorrelationsLsc}
\end{figure}

To better understand the dynamics of the anisotropic flow phase, in Fig. \ref{fig:CorrelationsLsc} we plot the velocity time correlation function
\begin{equation}
C_{vv}(\tau) = \left\langle \frac{\mathbf{v}(t + \tau) \cdot \mathbf{v}(t)}{|\mathbf{v}(t)|^2} \right\rangle 
\end{equation}
for simulations with $L_{SC} = 0.35$, $L_{SC} = 0.2$, and $L_{SC} = 0.1$. Interestingly, the dynamics differ depending on the boundary condition. Due to the overall circulation, the flows for spiral boundary conditions remain correlated for long times even as $L_{SC}$ is decreased. Near the transition ($L_{SC} = 0.35$), however, we find that homeotropic boundary conditions give correlated flows due to the residual circulation present in the system, while planar boundary conditions result in uncorrelated flows. As $L_{SC}$ decreases, systems with homeotropic boundary conditions become uncorrelated, while systems with planar anchoring become more correlated and require longer times to become uncorrelated. Additionally, the velocity correlation functions in the confined system are markedly different from those observed in unconfined systems, which exhibit completely uncorrelated flows at small $L_{SC}$ (Fig. S3).

The differences between the dynamics for planar and homeotropic boundary conditions can be explained by the average structure of the flows shown in Fig. \ref{fig:PerpVorticities}(b). For planar anchoring, the vortices on average form an azimuthal periodic structure around the boundary, while for homeotropic anchoring, all periodicity is destroyed as $L_{SC}$ diminishes and the vortices become smaller. These results suggest that both the structure and dynamics of the anisotropic flow phase may be tuned with the nematic boundary condition, which gives insight into how biological systems organize flows and has implications for technological applications of active fluids involving controlled mixing. 

{\it{Summary---}} In this work, using the hydrodynamic screening length as a control parameter, we show that circularly confined active nematics transition with decreasing screening length from a circulating flow phase to a previously undescribed anisotropic flow phase characterized by flow organized perpendicularly to the nematic boundary condition. Both dynamical phases feature organized flows distinct from those found in the well-known active turbulent phase. Our work shows that substrate friction and confinement can be used as control mechanisms for the directionality and dynamic correlations of flows via the nematic boundary conditions. While there has been some experimental work in which the substrate is altered such that an effective friction is varied \cite{guillamat16b,thijssen21}, the interplay between confinement and friction has been unexplored experimentally. Following the results of Ref. \cite{thijssen21}, we propose that the depth of the oil layer may be varied in a circularly confined microtubule based active nematic in order to reproduce our results experimentally. It has additionally been shown both experimentally and numerically that similar transitions occur in three-dimensional active nematics as the system becomes more confined \cite{wu17,chandragiri20,varghese20}. This indicates that three-dimensional confinement may act as an effective friction on the system and that the complex flows observed may potentially be explained by the simpler two-dimensional model used here.

Future work includes expanding the phase diagram for confined active nematics. In this study we have only varied the screening length $L_{SC}$, but we expect a rich dynamical phase landscape to emerge as the activity is also varied. This would lead to a better understanding of the interplay between the screening length, the nematic correlation length, and the active length. Additionally, different types of confinement may yield even more modes of control over active systems. We explored the effect of positive curvature, but negative curvature could be induced by a circular inclusion. Experiments in annuli have already shown controlled circulating behavior \cite{wu17,hardouin20} and immersed microstructures have been shown to pin defects \cite{figueroa22}. Due to the increasing degree of experimental and engineered control over boundary geometries and confinement, the understanding of how active fluids interact with their environment is becoming more important and practical.

\begin{acknowledgements}
This work was supported by the U.S. Department of Energy through the Los Alamos National Laboratory. Los Alamos National Laboratory is operated by Triad National Security, LLC, for the National Nuclear Security Administration of the U.S. Department of Energy (Contract No. 89233218CNA000001).
\end{acknowledgements}

\bibliography{LC}

\end{document}


\title{Supplemental Material for ``Friction mediated phase transition in confined active nematics''}

\author{Cody D. Schimming}
\email[]{cschim@lanl.gov}
\affiliation{Theoretical Division and Center for Nonlinear Studies, Los Alamos National Laboratory, Los Alamos, New Mexico, 87545, USA}

\author{C. J. O. Reichhardt}
\affiliation{Theoretical Division and Center for Nonlinear Studies, Los Alamos National Laboratory, Los Alamos, New Mexico, 87545, USA}

\author{C. Reichhardt}
\affiliation{Theoretical Division and Center for Nonlinear Studies, Los Alamos National Laboratory, Los Alamos, New Mexico, 87545, USA}

\maketitle

\section{Continuum active nematic model}
To simulate an active nematic we numerically solve a minimal continuum model for the time evolution of the tensor order parameter $\mathbf{Q}$ and the fluid velocity $\mathbf{v}$ \cite{marenduzzo07,doo18}:
\begin{gather}
\frac{\partial \mathbf{Q}}{\partial t} + \left( \mathbf{v} \cdot \nabla\right) \mathbf{Q} - \mathbf{S} = -\frac{1}{\gamma} \frac{\delta F}{\delta \mathbf{Q}} \label{eqn:Qevo} \\
-\eta \nabla^2 \mathbf{v} + \Gamma \mathbf{v} = -\nabla p - \alpha \nabla \cdot \mathbf{Q} \label{eqn:Stokes}\\
\nabla \cdot \mathbf{v} = 0 \label{eqn:Incompressible}
\end{gather}
where $F$ is the free energy of the nematic, $\gamma$ is a rotational viscosity, $\eta$ is the fluid viscosity, $\Gamma$ is the friction coefficient between the sample and substrate, $\alpha$ is the activity strength, and 
\begin{equation}
    \mathbf{S} = \left(\lambda\mathbf{E} + \bm{\Omega}\right)\left(\mathbf{Q} + \frac{1}{2}\mathbf{I}\right) + \left(\mathbf{Q} + \frac{1}{2}\mathbf{I}\right)\left(\lambda\mathbf{E} - \bm{\Omega}\right) - 2\lambda\left(\mathbf{Q} + \frac{1}{2}\mathbf{I}\right)\left(\nabla\mathbf{v} : \mathbf{Q}\right)
\end{equation}
is a generalized tensor advection\cite{beris94}. $2 \mathbf{E} = \nabla \mathbf{v} + \nabla \mathbf{v}^T$ is the rate of strain tensor, $2 \bm{\Omega} = \nabla \mathbf{v} - \nabla \mathbf{v}^T$ is the rotation tensor, and $\lambda$ is a dimensionless parameter that determines whether the liquid crystal is flow aligning or flow tumbling and depends on the geometry of the nematogen. The free energy we use is the two-dimensional Landau-de Gennes free energy:
\begin{equation}
    F = \int \left(A |\mathbf{Q}|^2 + C |\mathbf{Q}|^4 + L |\nabla \mathbf{Q}|^2\right) \, d\mathbf{r}
\end{equation}
where $A$ and $C > 0$ are phenomenological material parameters and $L$ is an elastic constant. This model has been shown to reproduce the flows and phenomenology of a diverse range of active nematic systems\cite{marchetti13,doo17,genkin17,doo18,sokolov19}.

We non-dimensionalize the system by working in dimensionless length and time,
\begin{equation}
    \mathbf{\tilde{r}} = \frac{\mathbf{r}}{\xi}, \quad \tilde{t} = \frac{t}{\sigma}, \quad \xi = \sqrt{\frac{\varepsilon L}{C}}, \quad \sigma = \frac{\gamma}{C \xi^2}
\end{equation}
where $\xi$ is the nematic correlation length, $\sigma$ is the nematic relaxation time, and $\varepsilon$ is a dimensionless parameter that controls the size of defects. The free energy is then written in units of $C\xi^2$ and Eq. \eqref{eqn:Stokes} may be multiplied by $\xi / C$ to be put in a non-dimensional form. This yields the dimensionless quantities
\begin{equation}
    \tilde{F} = \frac{F}{C\xi^2}, \quad a = \frac{A}{C}, \quad \tilde{p} = \frac{p}{C}, \quad \tilde{\eta} = \frac{\eta}{\sigma C}, \quad \tilde{\Gamma} = \frac{\Gamma \xi^2}{\sigma C}, \quad \tilde{\alpha} = \frac{\alpha}{C}.
\end{equation}

In confined active nematic systems, there are several length scales that affect the system dynamics. The nematic correlation length $\xi$ is the scale over which nematic distortions occur. In our system, all other lengths are compared to the correlation length. Another important length scale is the size of defects, which is controlled by the parameter $\varepsilon$ as noted above. The active length scale $L_{a} = \sqrt{L / \alpha}$ is the length at which active forces perturb the underlying nematic. It is correlated with the average distance between defects and hence inversely related to defect density. For confined systems, when $L_{a} \lessapprox R$ where $R$ is the system size, the active force is strong enough to unbind topological defects. In our work, we fix $L_{a}$, $\varepsilon$, and $R$. The other important length scale is the hydrodynamic screening length $L_{SC} = \sqrt{\eta / \Gamma}$, which is the primary focus of this work. This is the length scale below which hydrodynamic interactions are important.

There are also several time scales associated with the model. The nematic relaxation time $\sigma$ is the time scale over which elastic relaxation occurs in the nematic. We choose to compare all times in the system with the relaxation time. The time scale $\sigma_{a} = \gamma / \alpha$ is the time scale over which active stresses are passively relaxed. Similarly $\sigma_{\eta} = \eta / \alpha$ is the time scale over which active stresses are dissipated by viscous stresses and $\sigma_{\Gamma} = \xi^2 \Gamma / \alpha$ is the time scale over which active stresses are dissipated by frictional damping with the substrate. 

In our work, we fix the dimensionless ratio $\tilde{\alpha}/(\tilde{\eta} + \tilde{\Gamma})$. This fixes the ratio of timescales $(\sigma_{\eta} + \sigma_{\Gamma})/\sigma$ where the sum in the numerator is the average time scale of active stress injection \cite{giomi14}. This constraint differs from previous works that include frictional damping in which only $\tilde{\alpha}/\tilde{\eta}$ is held constant, thus leading to arrested flows as $\Gamma$, and, hence $\sigma_{\Gamma}$, is increased \cite{doo16,doo18}. This can be seen explicitly by comparing the resulting Stokes equations. If $\tilde{\alpha}/(\tilde{\eta} + \tilde{\Gamma}) = 1$ is fixed, Eq. \eqref{eqn:Stokes} may be written (after rescaling the pressure)
\begin{equation}
    -\tilde{L}_{SC}^2 \nabla^2 \mathbf{v} + \mathbf{v} = -\nabla p - (1 + \tilde{L}_{SC}^2)\nabla \cdot \mathbf{Q}.
\end{equation}
In this case the active force still plays a role in both limits of large and small $\tilde{L}_{SC}$ since it is never dominated by the terms on the left hand side of the equation. On the other hand, if only $\tilde{\alpha}/\tilde{\eta} = 1$ is fixed Eq. \eqref{eqn:Stokes} becomes
\begin{equation}
    -\tilde{L}_{SC}^2 \nabla^2 \mathbf{v} + \mathbf{v} = -\nabla p - \tilde{L}_{SC}^2\nabla \cdot \mathbf{Q}.
\end{equation}
Here, in the small $\tilde{L}_{SC}$ limit (i.e. the high friction limit) the damping term proportional to the velocity dominates all other terms and the resulting velocity goes to zero. Thus, our constraint allows us to vary the screening length without damping the active dynamics. In the main text, and in following sections, the tildes denoting dimensionless quantities are omitted for brevity.

\section{Numerical Implementation}
To numerically solve Eqs. \eqref{eqn:Qevo}, \eqref{eqn:Stokes}, and \eqref{eqn:Incompressible}{\color{red},} we first write $\mathbf{Q}$ in a basis for two dimensional traceless, symmetric tensors
\begin{equation}
    \mathbf{Q} = \begin{pmatrix}
                 q_1 & q_2 \\
                 q_2 & -q_1
                 \end{pmatrix}
\end{equation}
and rewrite Eq. \eqref{eqn:Qevo} in terms of $q_1$ and $q_2$. To solve Eq. \eqref{eqn:Stokes} we use the stream function formulation and write $\mathbf{v} = \left(\partial_y \psi,-\partial_x \psi\right)$. Using this, the pressure term drops out of Eq. \eqref{eqn:Stokes} and Eq. \eqref{eqn:Incompressible} is automatically satisfied. The full equations for $\psi$, $q_1$, and $q_2$, are
\begin{gather}
    \eta \nabla^4 \psi - \Gamma \nabla^2 \psi = \alpha \left(2\partial_{xy} q_1 - \partial_{xx}q_2 + \partial_{yy} q_2\right) \label{eqn:psi} \\
    \begin{split}
    \partial_t q_1 + \left(\mathbf{v}\cdot\nabla\right) q_1 - \lambda\left[\partial_{xy}\psi - 4 q_1^2 \partial_{xy}\psi - 2 q_1 q_2\left(-\partial_{xx}\psi + \partial_{yy}\psi\right)\right] - q_2\nabla^2\psi \\ = -4 a q_1 - 16 q_1\left(q_1^2 + q_2^2\right) + \frac{1}{\varepsilon}\nabla^2 q_1 
    \end{split} \label{eqn:q1} \\
    \begin{split}
    \partial_t q_2 + \left(\mathbf{v}\cdot\nabla\right) q_2 - \lambda\left[\frac{1}{2}\left(-\partial_{xx}\psi + \partial_{yy}\psi\right) - 4 q_1 q_2\partial_{xy}\psi - 2q_2^2\left(-\partial_{xx}\psi + \partial_{yy}\psi\right)\right] + q_1\nabla^2 \psi \\ = -4 a q_2 - 16 q_2\left(q_1^2 + q_2^2\right) + \frac{1}{\varepsilon}\nabla^2 q_2.
    \end{split} \label{eqn:q2}
\end{gather}

Equations \eqref{eqn:psi}, \eqref{eqn:q1}, and \eqref{eqn:q2} are discretized in space in a circular domain of radius $7.5$ on a triangular mesh using the Matlab/C++ package FELICITY \cite{walker18}. Equations \eqref{eqn:q1} and \eqref{eqn:q2} are discretized in time using a backward-Euler method with time step $\delta t = 0.5$. We set $a = -0.5$ so the non-active liquid crystal is in the nematic phase and $\varepsilon = 4$ so defects are of roughly unit length. We also set $\lambda = 1$ so the nematic is flow aligning and we determined the Leslie angle to be $\theta_{L} \approx 0.33$ \cite{leslie92}. We maintain $\alpha = 1$ and the sum of the dimensionless parameters $\eta + \Gamma = 1$. The system is initialized with a random director field except at the fixed boundaries. At each time step, Eq. \eqref{eqn:psi} is first solved to yield the velocity field of the given nematic configuration. The discretized, nonlinear equations \eqref{eqn:q1} and \eqref{eqn:q2} are then solved using a Newton-Raphson method. To fix the anchoring conditions of the nematic, we employ Dirichlet conditions on the order parameter at the boundary. This is repeated for $1000$ time steps.

\section{Effect of hydrodynamic screening on unconfined active nematics}
Here we describe the effect of decreasing $L_{SC}$ on unconfined active nematics with periodic boundary conditions. Our approach differs from previous investigations of the effect of friction on bulk systems in that we fix the value of $\alpha / (\eta + \Gamma)$. Thus, to properly compare the confined case to unconfined active nematics using a similar methodology, we also simulate active nematics on a square domain with periodic boundary conditions. We set the parameters to be the same as above and fix the domain size to $[-7.5,7.5]^2$.

Using our method of reducing $L_{SC}$ we find a transition between a periodic flow state and a traditional active turbulence state. We plot in Fig.~\ref{fig:BulkVorticities}(a) a time snapshot of the velocity and vorticity fields in the active turbulent phase, while in Fig.~\ref{fig:BulkVorticities}(b) we plot the time average of the velocity and vorticity fields over the whole simulation. Since the system is unconfined, the many vortices that appear in this phase have no structural organization and hence, time average to zero as shown in Fig.~\ref{fig:BulkVorticities}(b). To quantify this transition we measure the velocity time correlation function
\begin{equation}
    C_{vv}(\tau) = \left\langle\frac{\mathbf{v}(t + \tau)\cdot\mathbf{v}(t)}{|\mathbf{v}(t)|^2}\right\rangle
\end{equation}
and extract the characteristic timescale of decoherence, $\tau^*$. Figure \ref{fig:BulkCoherences}(a) shows $\tau^*$ plotted against a range of $L_{SC}$. At large $L_{SC}$ the flow is periodic, as shown in Fig. \ref{fig:BulkCoherences}(b) where we plot $C_{vv}(\tau)$ for an unconfined system with $L_{SC} = 50$. Here the decoherence time is long but still finite. When $L_{SC} \sim 1$ the system transitions to active turbulence. Here, the decoherence times are very short and the dynamics has no correlations as shown in Fig.~\ref{fig:BulkCoherences}(c). 

The transition to active turbulence may be understood from the fact that $L_{SC}$ sets the size of vorticies. When $L_{SC}$ is much larger than the periodic system, a vortex cannot materialize since it does not fit inside the domain, and the flow instead organizes along lanes. These flow lanes change the orientation of the nematic, causing flow to develop in the perpendicular direction, and the process repeats, creating the periodic dynamics observed. As $L_{SC}$ decreases, the characteristic vortex size decreases and several vortices may form in the system. If the vortices are on the scale of the system size they persist, leading to longer decoherence times. Note that the longest decoherence times shown in Fig. \ref{fig:BulkCoherences}(a) are for $L_{SC} \sim 15$ which is precisely the simulated system size. When $L_{SC}$ is much smaller than the system size, many vortices form and the typical active turbulent behavior dominates. These vortices tend to be long and thin like those found in the confined system, but because there is no fixed director to guide their organization, they are oriented randomly.

\section{Supplemental Movies}
\begin{itemize}
    \item Supplemental Movie 1: Nematic configuration (left) and velocity and vorticity fields (right) as a function of time for a simulated active nematic with planar anchoring and $L_{SC} = 0.2$.
    \item Supplemental Movie 2: Nematic configuration (left) and velocity and vorticity fields (right) as a function of time for a simulated active nematic with homeotropic anchoring and $L_{SC} = 0.2$.
    \item Supplemental Movie 3: Nematic configuration (left) and velocity and vorticity fields (right) as a function of time for a simulated active nematic with spiral anchoring and $L_{SC} = 0.2$.
\end{itemize}

\section{Supplemental Figures}

\renewcommand{\thefigure}{S1}
\begin{figure}[h]
    \centering
    \includegraphics[width = 0.9\textwidth]{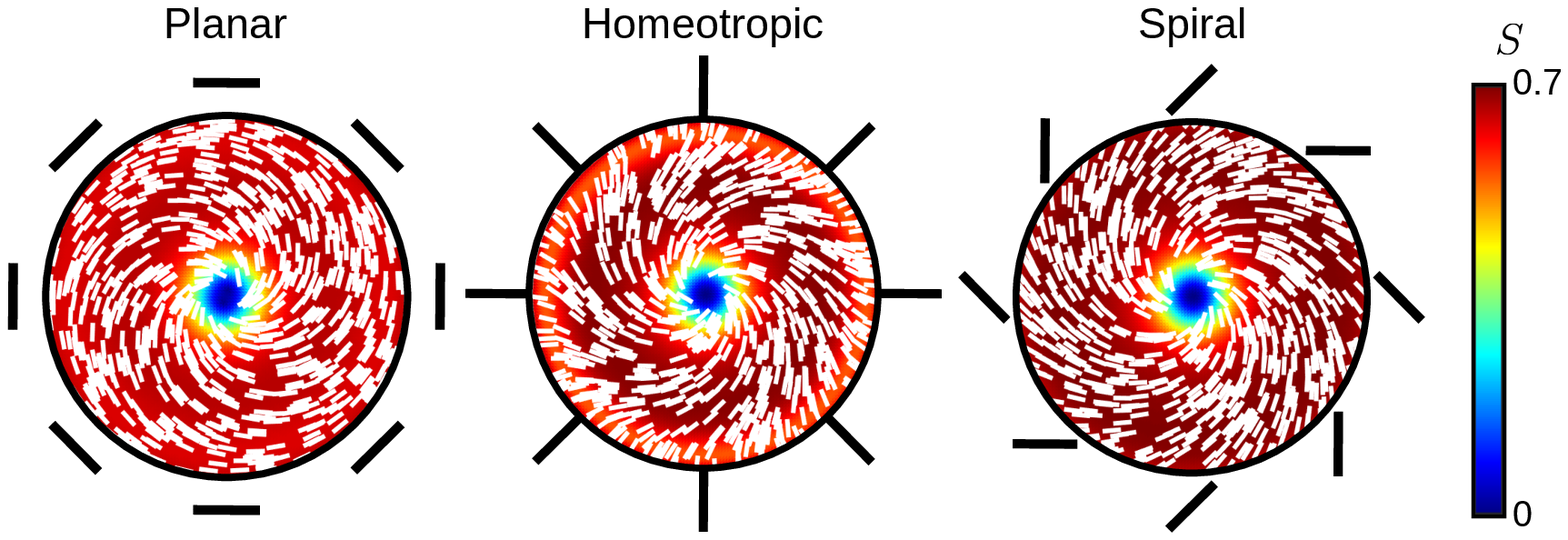}
    \caption{Nematic configuration for simulations with $L_{SC} = 10$ for planar, homeotropic, and spiral anchoring. In all cases, a central topological defect of charge $+1$ is formed with a local spiral pattern.}
    \label{fig:CirculationNematic}
\end{figure}

\renewcommand{\thefigure}{S2}
\begin{figure}[h]
    \centering
    \includegraphics[width = 0.9\textwidth]{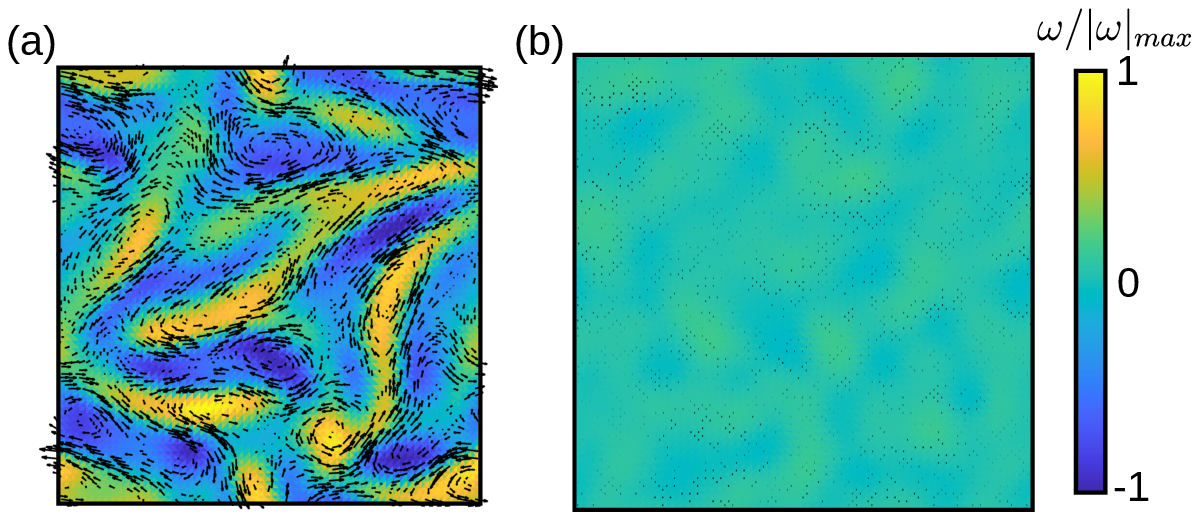}
    \caption{(a) Snapshot of the velocity (lines) and vorticity (color) fields for an unconfined active nematic simulation with periodic boundary conditions and $L_{SC} = 2$. (b) Time-averaged velocity and vorticity fields for the same simulation as in (a). In the unconfined system there is no overall structure and the velocity and vorticity time-average to zero.}
    \label{fig:BulkVorticities}
\end{figure}

\renewcommand{\thefigure}{S3}
\begin{figure}[h]
    \centering
    \includegraphics[width = \textwidth]{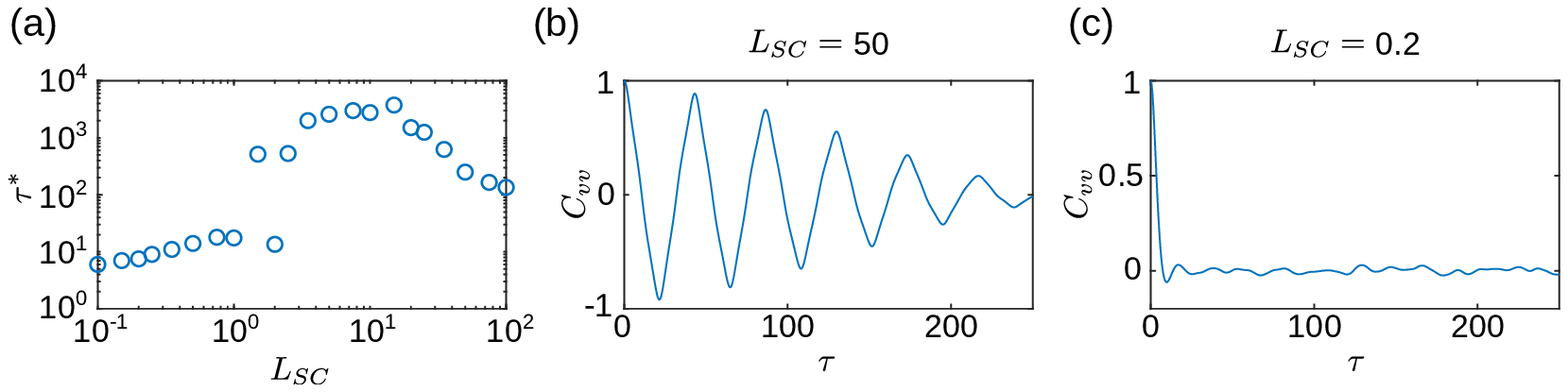}
    \caption{(a) Decoherence time $\tau^*$ in unconfined active nematics plotted against hydrodynamic screening $L_{SC}$. (b) Velocity time correlation function $C_{vv}$ versus advanced time $\tau$ for an unconfined active nematic simulation with $L_{SC} = 50$. The dynamics are periodic, but become uncorrelated after long times. (c) $C_{vv}$ versus $\tau$ for an unconfined active nematic with $L_{SC} = 0.2$. Here, the system is turbulent with very short decoherence times.}
    \label{fig:BulkCoherences}
\end{figure}

\newpage

\bibliography{LC}